\documentclass[12pt]{article}
\usepackage{graphicx}
\usepackage{subfigure}

\newcommand\pubnumber{SNSN-323-63}
\newcommand\pubdate{\today}
\def\institute{Institut f{\"u}r Physik\\
	Humboldt-Universit{\"a}t zu Berlin, Germany}
\def\support{\footnote{Copyright 2018 CERN for the benefit of the ATLAS Collaboration. Reproduction of this article or parts of it is allowed as specified in the CC-BY-4.0 license.}}

\def\Title#1{\begin{center} {\Large #1 } \end{center}}
\def\Author#1{\begin{center}{ \sc #1} \end{center}}
\def\Address#1{\begin{center}{ \it #1} \end{center}}

\newcommand\pubblock{\rightline{\begin{tabular}{l} \pubnumber\\
         \pubdate  \end{tabular}}}
\newenvironment{Abstract}{\begin{quotation}  }{\end{quotation}}
\newenvironment{Presented}{\begin{quotation} \begin{center} 
             PRESENTED AT\end{center}\bigskip 
      \begin{center}\begin{large}}{\end{large}\end{center} \end{quotation}}

\def\beq{\begin{equation}}
\def\eeq#1{\label{#1}\end{equation}}
\def\eeqn{\end{equation}}

\def\beqa{\begin{eqnarray}}
\def\eeqa#1{\label{#1}\end{eqnarray}}
\def\eeqan{\end{eqnarray}}

\let\bar=\overbar

\def\Dslash{\not{\hbox{\kern-4pt $D$}}}
\def\dslash{\not{\hbox{\kern-2pt $\del$}}}

\def\msb{{\bar{\ssstyle M \kern -1pt S}}}

\begin{document}
\begin{titlepage}
\pubblock

\vfill
\Title{Measurement of differential $t$­-channel single top-­quark production cross-sections with ATLAS}
\vfill
\Author{ Pienpen Seema \\
	on behalf of the ATLAS collaboration\support}
\Address{\institute}
\vfill
\begin{Abstract}
Absolute and normalised differential cross-sections of single top quarks produced in the $t$-channel are presented. 20.2~fb$^{-1}$ of proton-proton collision data at a centre­-of-­mass energy of 8~TeV collected by the ATLAS experiment at the LHC are used. Differential cross-sections as a function of the transverse momentum and the absolute value of the rapidity of the top quarks and the top antiquarks are measured at both parton level and particle level. The transverse momentum and rapidity differential cross-sections of the scattered light-quark jets are extracted at particle level. The measured cross-sections are compared to various Monte Carlo predictions as well as to available fixed-order QCD calculations. All results agree with the Standard Model predictions.
 
\end{Abstract}
\vfill
\begin{Presented}
$11^\mathrm{th}$ International Workshop on Top Quark Physics\\
Bad Neuenahr, Germany, September 16--21, 2018
\end{Presented}
\vfill
\end{titlepage}
\def\thefootnote{\fnsymbol{footnote}}
\setcounter{footnote}{0}

\section{Introduction}
At the LHC, top quarks can be produced singly via electroweak interactions. In leading-order perturbative QCD theory, there are three different single top-quark production mechanisms. They are distinguished according to the virtuality of the exchanged $W$ boson. The dominant process is the $t$-channel production where a light quark interacts with a bottom quark by exchanging a space-like $W$ boson. The other two processes are the $s$-channel in which a time-like $W$ boson is exchanged and the $Wt$-channel where a top quark is produced in association with a $W$ boson. At next-to-leading order (NLO) in perturbative QCD, the total cross-sections of top-quark and top-antiquark production in the $t$-channel at $\sqrt{s} = 8~\textrm{TeV}$, calculated with \textsc{HatHor}~v2.1~\cite{Kant:2014oha}, are predicted to be:  
\begin{eqnarray*}
	\sigma(tq)       & = & 54.9 ^{+2.3}_{-1.9}\ \mathrm{pb}, \\
	\sigma(\bar{t}q) & = & 29.7 ^{+1.7}_{-1.5}\ \mathrm{pb}, \\
	\sigma(tq+\bar{t}q) & = & 84.6 ^{+3.9}_{-3.4}\ \mathrm{pb}.
\end{eqnarray*}

\section{Measurement}
In this analysis~\cite{paper}, $pp$ collisions at $\sqrt{s} = 8~\textrm{TeV}$, corresponding to an integrated luminosity of 20.2~fb$^{-1}$, recorded with the ATLAS detector~\cite{PERF-2007-01} at the LHC are used. The signal event signature contains exactly one charged lepton (either $e$ or $\mu$), exactly two jets, one of the two jets being a $b$-tagged jet, and missing transverse momentum. One special characteristic of the $t$-channel topology is that the light-quark jet tends to be scattered in the forward direction. $t$-channel events are simulated using \textsc{Powheg-Box}~\cite{Frederix:2012dh} generator with the \textsc{Pythia6}~\cite{Sjostrand:2006za} for modelling of the parton shower, the hadronisation and the underlying event.

The parton-level measurements are based on the top quarks, $t$, over the full kinematic range. These top quarks are defined as the top quarks before their decay but after gluon radiation. This makes comparisons to theoretical predictions straightforward. For the particle-level measurements, the top quarks are reconstructed from stable particles in a fiducial phase space defined close to the experimental phase space. The stable particles are defined as having a mean lifetime greater than $0.3\times10^{-10}$~s. These top quarks are called pseudo-top-quarks~\cite{Aad:2015eia}, $\hat{t}$. Similarly, a scattered light-quark jet at particle level is written as $\hat{j}$. Both absolute and normalised differential cross-sections are measured in bins of $p_{\rm{T}}(t)$ and $|y(t)|$ at parton level as well as of $p_{\rm{T}}(\hat{t})$, $|y(\hat{t})|$, $p_{\rm{T}}(\hat{j})$ and $|y(\hat{j})|$ at particle level. 

A neural network (NN) is utilised to discriminate signal from background events. Good separation is achieved by using seven input variables where $|\eta{(j)}|$ is sorted as the second most powerful variable. This NN is used for all measurements except for the measurement as a function of $|y(\hat{j})|$ where a second NN is trained without $|\eta{(j)}|$ to avoid a distortion of the $|y(j)|$ distribution due to a cut on the NN output. All differential cross-sections are extracted in a signal enriched region where a cut on the NN output is applied. In this region, the signal-over-background ratio of about 2 is achieved.
		
As the measured distributions are distorted by detector effects and selection efficiency, they are unfolded using D'Agostini's iterative approach~\cite{DAgostini1995487}, implemented inside RooUnfold framework~\cite{Adye:2011gm}, in order to recover their true distribution. This leads to direct comparisons between the unfolded distributions and theoretical predictions. Several tests are performed to ensure that the chosen unfolding method is reliable and can be used for the analysis.

\section{Results}
\begin{figure}[htbp]
	\centering
	\subfigure[]{
		\includegraphics[trim={0 18pt 0 37pt}, clip, width=0.45\textwidth]{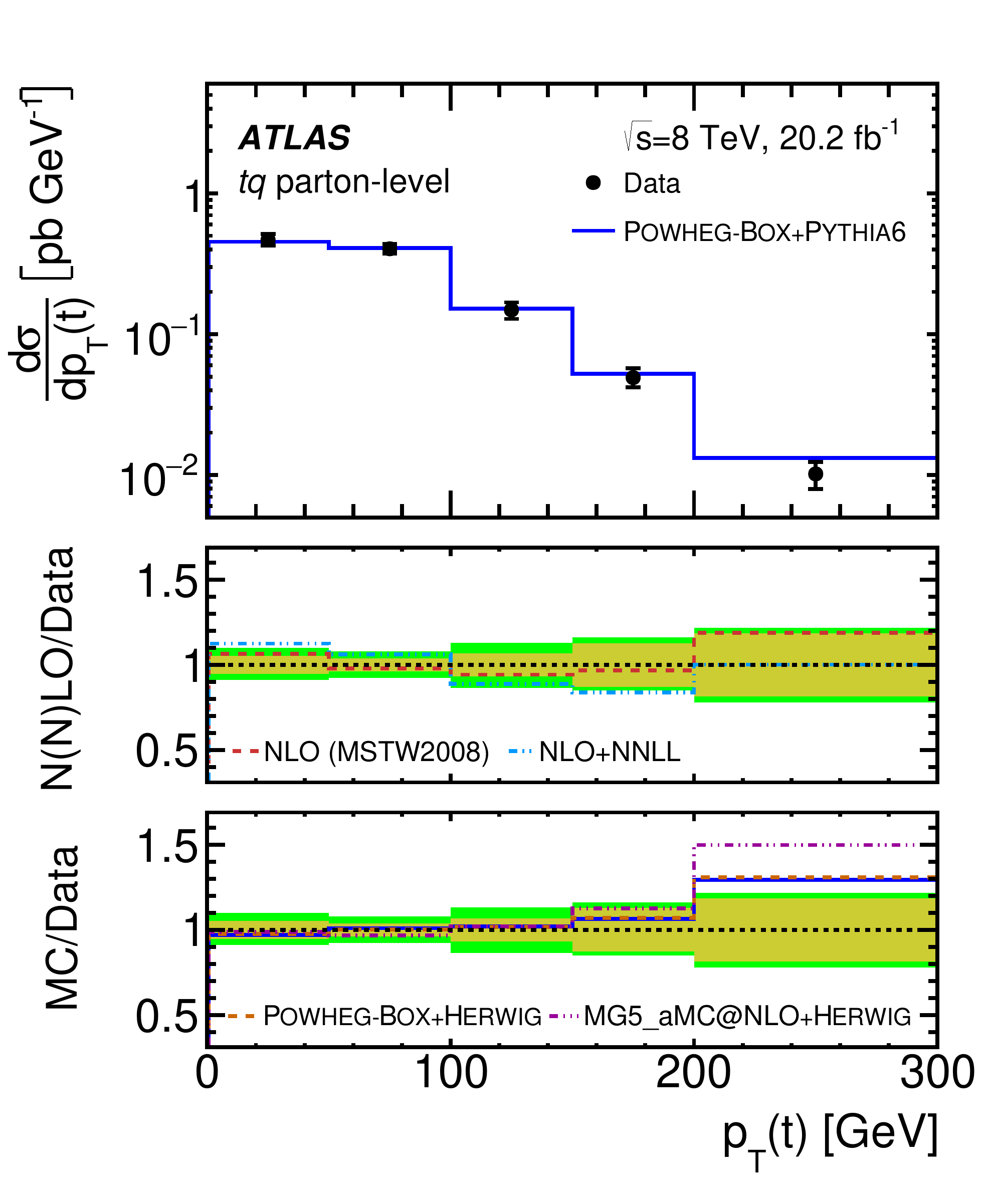}
	}
	\subfigure[]{
		\includegraphics[trim={0 18pt 0 37pt}, clip, width=0.45\textwidth]{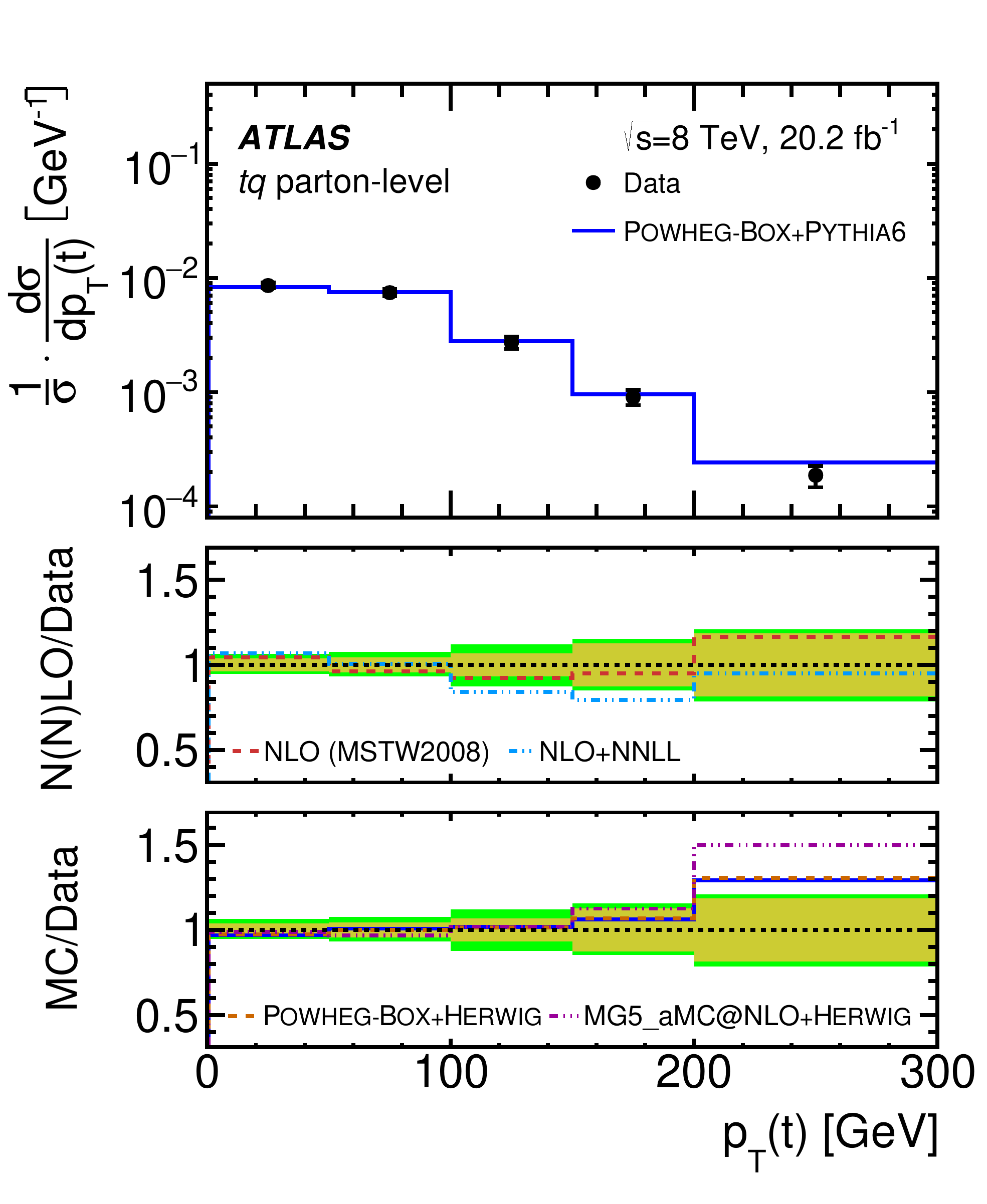}
	}
	\subfigure[]{
		\includegraphics[trim={0 18pt 0 37pt}, clip, width=0.45\textwidth]{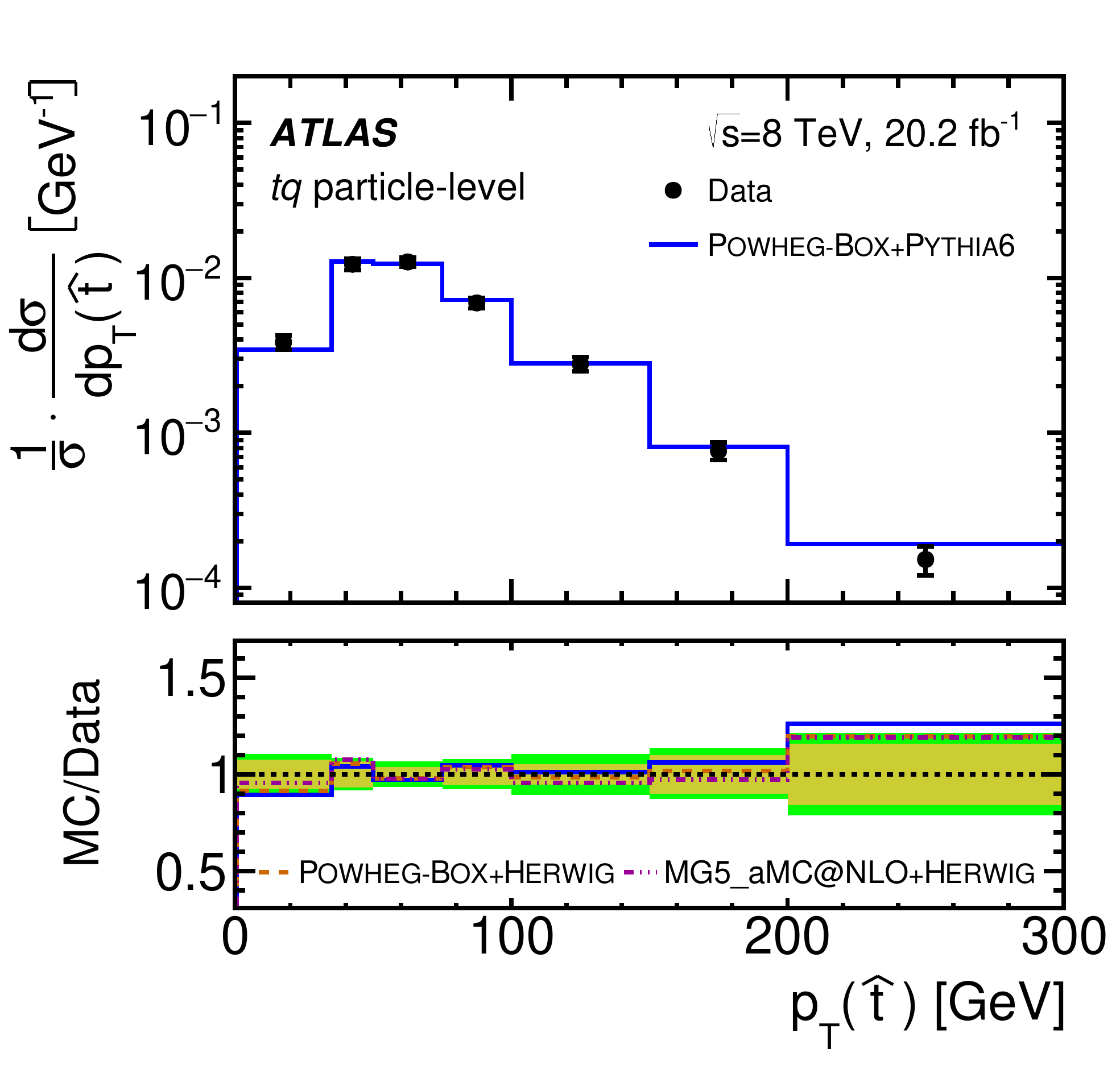}
	}
	\subfigure[]{
		\includegraphics[trim={0 18pt 0 37pt}, clip, width=0.45\textwidth]{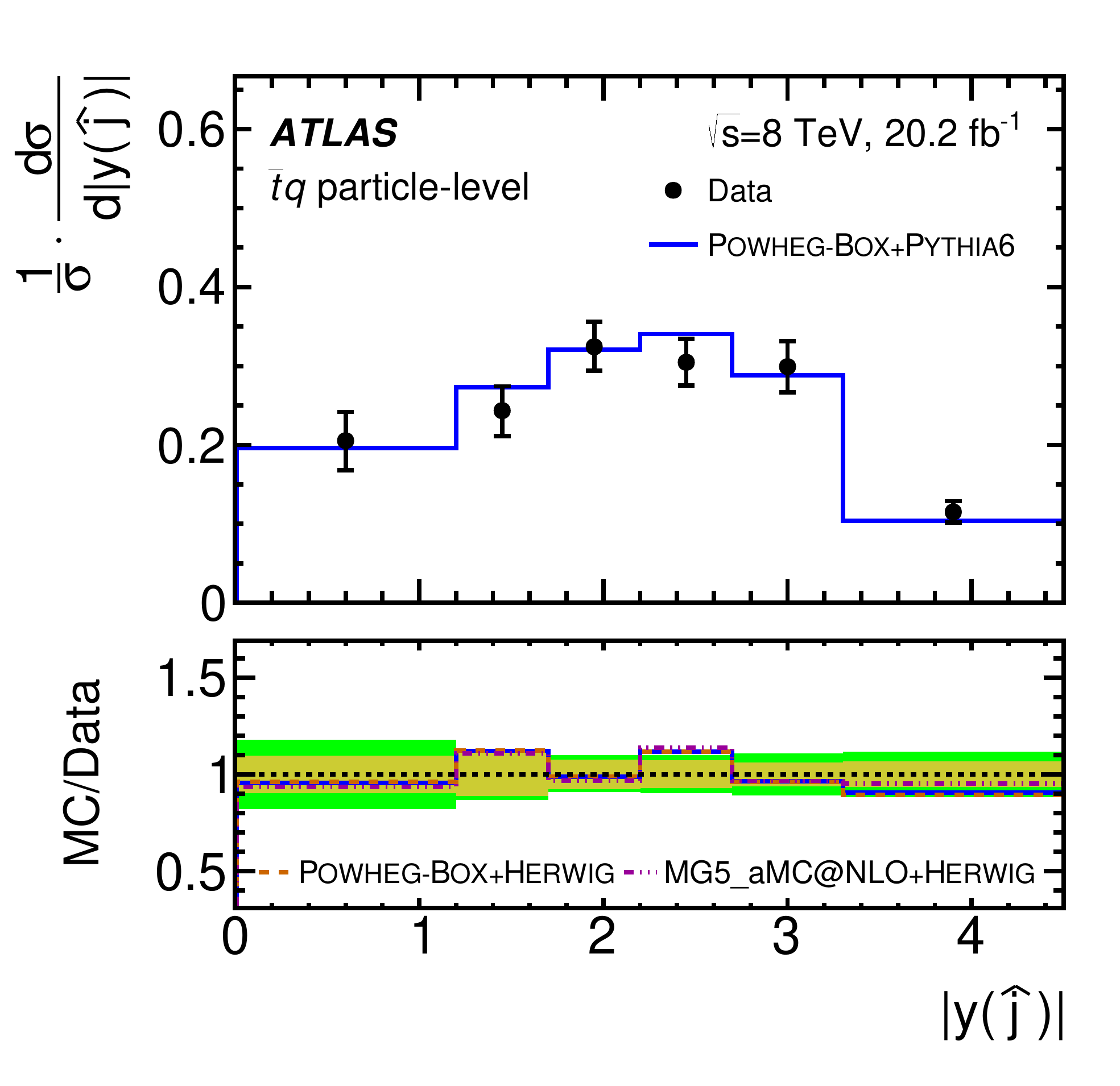}
	}	
	\caption{(a) absolute and (b) normalised differential cross-sections for $tq$ production at parton level as a function of $p_{\rm{T}}(t)$. Particle-level normalised differential cross-sections (c) for $tq$ production as a function of $p_{\rm{T}}(\hat{t})$ as well as (d) for $\bar{t}q$ production as a function of $|y(\hat{j})|$. The distributions are compared to various MC predictions (and available fixed-order QCD calculations). The vertical error bars on the data points represent the total uncertainty. In the bottom part of each figure, the inner (outer) error band represents the statistical (total) uncertainty.~\cite{paper}}
	\label{fig:ptcl}
\end{figure}

A selection of the results for the absolute and normalised unfolded differential cross-sections for $tq$ and $\bar{t}q$ production is shown in Figure~\ref{fig:ptcl}.

The parton-level cross-sections are compared to different MC predictions using the \textsc{Powheg-Box} and \textsc{Madgraph5}\_aMC@NLO~\cite{Alwall:2014hca} generators. Comparisons to different parton-shower and hadronisation models are shown: \textsc{Pythia6} or \textsc{Herwig}~\cite{Corcella:2000bw} interfaced to \textsc{Powheg-Box}. It can be observed that the parton-shower and hadronisation modelling has a very small effect on the predictions. The measured cross-sections are also confronted with NLO QCD predictions calculated using MCFM~\cite{Campbell}. A calculation at approximate NLO+NNLL QCD is available for the top-quark and top-antiquark $p_{\rm{T}}$ distributions~\cite{Kidonakis:2013yoa}. Good agreement is seen between the data and all predictions, with the same tendency for almost all MC predictions to be somewhat harder than the data as a function of $p_{\rm{T}}(t)$. The data is described better by the approximate NLO+NNLL prediction than the MC predictions as a function of $p_{\rm{T}}(t)$.

The particle-level cross-sections are compared to the same MC predictions used for the comparison of the parton-level cross-sections. All measured differential cross-sections agree well with the MC predictions.
More bins can be used at particle level, due to better resolution of the pseudo-top-quarks.

The measurements are at $5-20\%$ precision per bin. The major sources of uncertainty are the uncertainties associated with the modelling of the signal and the top-background processes as well as the uncertainties in the jet energy scale calibration. The uncertainties due to the unfolding procedure are small compared to the total uncertainty. In general, the total systematic uncertainty for the normalised differential cross-sections is smaller than the uncertainty for the absolute differential cross-sections because many systematic uncertainties are reduced or cancelled for the normalised cross-section measurements.

\section{Conclusion}
Measurements of absolute and normalised differential cross-sections for $tq$ and $\bar{t}q$ production with the ATLAS detector using 20.2~fb$^{-1}$ of $pp$ collision data at $\sqrt{s}~=~8~\textrm{TeV}$ are presented. Selected events are characterised by one electron or muon, missing transverse momentum, and two jets. One of the jets must be a $b$-tagged jet. Differential cross-sections as a function of the transverse momentum and the absolute value of the rapidity of the top (anti)quarks are extracted over the full kinematic range at parton-level. Differential cross-sections as a function of the transverse momentum and the absolute value of the rapidity of the top (anti)quarks and the scattered light-quark jets are measured within a fiducial phase space at particle level for the first time. The most precise measurements are the normalised differential cross-sections at particle level. All measured differential cross-sections confirm the Standard Model predictions. No indication of new physics is observed in the presented results.


\begin{thebibliography}{99}


\bibitem{Kant:2014oha}
P. Kant et al., Comput. Phys. Commun. {\bf 191} (2015), 74. 

\bibitem{paper}
ATLAS Collaboration, Eur. Phys. J., {\bf C77} (2017), 531.

\bibitem{PERF-2007-01}
ATLAS Collaboration, JINST, {\bf 3} (2008), S08003. 

\bibitem{Frederix:2012dh}
R. Frederix et al., JHEP, {\bf 09} (2012), 130. 

\bibitem{Sjostrand:2006za}
T. Sj{\"o}strand et al., JHEP, {\bf 05} (2006), 026. 

\bibitem{Aad:2015eia}
ATLAS Collaboration, JHEP, {\bf 06} (2015), 100.

\bibitem{DAgostini1995487}
G. D'Agostini, Nucl. Instrum. Meth. A, {\bf 362} (1995), 487. 

\bibitem{Adye:2011gm}
T. Adye, arXiv:1105.1160.

\bibitem{Alwall:2014hca}
J. Alwall et al., JHEP, {\bf 07} (2014), 079. 

\bibitem{Corcella:2000bw}
G. Corcella et al., JHEP, {\bf 01} (2001), 010. 

\bibitem{Campbell}
J. M. Campbell et al., Phys. Rev. D, {\bf 70} (2004), 094012. 

\bibitem{Kidonakis:2013yoa}
N. Kidonakis, Phys. Rev. D, {\bf 88} (2013), 031504. 






\end{thebibliography}
\end{document}